\input harvmac 





\Title{\vbox{\hsize=3.truecm \hbox{SPhT/02-187}}}
{{\vbox {
\bigskip
\centerline{A Comment on ``Free energy fluctuations in Ising}
\centerline{spin glasses", by T. Aspelmeier and M.A. Moore}
}}}
\bigskip
\centerline{C. De Dominicis\foot{cirano@spht.saclay.cea.fr}  and
P. Di Francesco\foot{philippe@spht.saclay.cea.fr}}
\medskip
\bigskip
\centerline{ \it Service de Physique Th\'eorique, CEA/DSM/SPhT}
\centerline{ \it Unit\'e de recherche associ\'ee au CNRS}
\centerline{ \it CEA/Saclay}
\centerline{ \it 91191 Gif sur Yvette Cedex, France}
\bigskip
\vskip .5in
\noindent We show that there is no need to modify the Parisi replica
symmetry breaking ansatz, by working with $R$ steps of breaking and solving
{\it exactly} the discrete stationarity equations generated by the 
standard ``truncated Hamiltonian" of spin glass theory.  
\Date{01/03}

\nref\AMOOR{T. Aspelmeier and M.A. Moore, {\it Free energy fluctuations in 
Ising spin glasses}, preprint cond-mat/0211707 (2002), hereafter
referred as AM.} 
\nref\PAR{G. Parisi, Phys. Lett. {\bf 73A} (1979) 203 and
J. Phys. {\bf A13} (1980) L115 and 1101.}
\nref\KKK{I. Kondor, J. Phys. {\bf A16} (1983) L127.}
\nref\DD{C. De Dominicis and P. Di Francesco, in preparation.}
\nref\DCT{C. De Dominicis, D.M. Carlucci and T. Temesvari,
J. Phys. I France {\bf 7} (1997) 105. In the continuum limit,
the transform was first used in M. Mezard and G. Parisi, J. Phys. I
France {\bf 1} (1991) 809.}
\nref\DKT{C. De Dominicis, I. Kondor and T. Temesvari, in ``Spin
glasses and random fields", edited by A.P. Young (World Scientific,
Singapore, London 1997). We keep here the same notations except
for multiplicities in which we exhibit their factor $n$.}
\nref\AMYSIX{T. Aspelmeier, M.A. Moore and A.P. Young, {\it
Interface energies in Ising spin glasses}, preprint
cond-mat/0209290 (v1) (2002).}
\nref\AMYHUIT{T. Aspelmeier, M.A. Moore and A.P. Young, {\it
Interface energies in Ising spin glasses}, preprint
cond-mat/0209290 (v2) (2002).}

\newsec{Introduction}

In a quite recent work Aspelmeier and Moore \AMOOR\ have considered
the sample-to-sample free energy fluctuations in finite dimensional
spin glasses via the replica method. To that effect they reconsider 
higher order terms in the replica number $n$ and they conclude that
the Parisi symmetry breaking scheme \PAR\ does not give the correct answer
for these higher order terms. Finally they propose a modified 
symmetry breaking scheme that resolves the problem.

What we set out to do here is the following.
Starting from the same truncated Hamiltonian (AM.3) {\it we solve
exactly the discrete stationarity equations} for $R$ steps of replica
symmetry breaking, namely we obtain the $R+1$ values of $q_{\alpha\beta}$
indexed by their overlap values $q_0,q_1,...,q_R$
(together with $q_{\alpha\alpha}\equiv q_{R+1}=0$) and the $R$ values
of Parisi box sizes $p_1,p_2,...,p_R$ together with the two {\it fixed}
boundary values $p_0=n$ and $P_{R+1}=1$.
As a result, we find two families (a), (b) of solutions associated with two
possible values of $q_0$, namely, letting $g=w/(2 y)$ 
\item{(a)} $q_0= {3n\over 2} g$ 
\par
In this case the corresponding free energy is identical to the Kondor \KKK\ 
result
\eqn\konres{ n f^{(a)}(n)= n f -{9 n^6\over 640} w g^3 }
\item{(b)} $q_0=0$
\par
The free energy is now larger
\eqn\frehi{ n f^{(b)}(n)= n f }

The (b) solution is therefore the appropriate one to choose, both solutions 
having a non-negative Hessian spectrum when $R\to \infty$.
Among the family of solutions (b) with $q_0=0$ and free energy $f^{(b)}$,
we will pick a {\it reference} solution with a set of values $q_t,p_t$, 
$t=1,2,...,R$. All the other solutions will be shown elsewhere \DD\ 
to correspond to a (discrete) reparametrization for large $R$. With that set
of values, we proceed and compute the contribution to fluctuations, with
a result that matches for $R\to \infty$ the Aspelmeier and Moore ones \AMOOR.
We thereby establish that there is indeed no need for modifying the Parisi
replica symmetry breaking scheme. 

\newsec{Solution of the stationarity equations}

The stationarity equations are derived from the free energy functional 
\eqn\frefunc{n f= -\sum_{t=0}^{R+1} \left\{ (p_t-p_{t+1})\big[ {\tau \over 2} q_t^2 
+{u\over 12} q_t^4\big] +\big({1\over p_t}-{1\over p_{t-1}}\big) {w\over 6}
{\hat q}_t^3 \right\} }
where we have used the replica Fourier transform $\hat q$ of $q$ \DCT\
\eqn\four{ {\hat q}_k=\sum_{t=k}^{R+1} p_t(q_t-q_{t-1})=\sum_{t=k}^R p_t(q_t-q_{t-1})-q_R}
Combining the stationarity equations, we obtain in the end
\eqn\soldisc{\eqalign{ g p_t &= {1\over 2}(q_t+q_{t-1}) \qquad t=1,2,...,R\cr
(q_t-q_{t-1})^2&=(q_{t-1}-q_{t-2})^2=...=(q_1-q_0)^2\qquad t=1,2,...,R\cr}}
Here we concentrate on the particular {\it reference}
solution such that 
\eqn\refesol{ q_t-q_{t-1}=q_{t-1}-q_{t-2}=...=q_1-q_0={q_r-q_0\over R} }
which leads to
\eqn\exact{\eqalign{
q_t&=q_0 +(q_R-q_0) {t\over R} \qquad t=0,1,...,R\cr
g p_t&= q_0+(q_R-q_0) {2t-1 \over 2R} \qquad  t=1,2,...,R\cr}}
together with $q_{R+1}=0$, $p_{R+1}=1$.
Besides one has two more equations that determine $q_0$ and $q_R$
\eqn\oneq{ E(q_R)-{u\over 6} \left({q_R-q_0\over R}\right)^2 =0}
where $E(q_R)=\tau -w q_R +u q_R^2$, and which is valid for $R>0$, and
\eqn\twoeq{ q_0\big(E(q_R)+{q_0\over 3}(3 g p_0-2 q_0) \big)=0}
valid for all $R$. Note that if $R=0$, $q_R\equiv q_0$, then \exact\
is a tautology and only \twoeq\ survives, leading to the standard
result $w q=2\tau/(2-p_0) +O(\tau^2)$. In fact, one is interested in the
limit of large $R$, whereby \oneq\ yields the relationship
\eqn\relshi{ E(q_R)=0}
and from \twoeq\ either $q_0=0$ or $q_0=3g p_0/2$ as respectively
in the cases (b) and (a). 
Note that $q_t$ is monotonous except for its last step ($q_{R+1}=0$), and
$p_t$ is monotonous except for its first step (when $p_0$ is kept fixed
at a value $n\neq 0$). 

In the {\it continuum limit}, where $t/R\to x$ and $q_t\to q(x)$,
$p_t\to p(x)$, we get for $x$ in the
{\it open} interval $(0,1)$ 
\eqn\qofx{ q(x)= g p(x) = g q_R x \qquad 0<x<1}
 
We now proceed to get the fluctuation contribution as in (AM.5).

\newsec{Fluctuations: the Replicon sector}

We have as in (AM.9)
\eqn\fren{ n \delta f_{Rep} ={V\over 2} \int {d^Dp\over (2\pi)^D} I_{Rep}(p)}
where
\eqn\irep{ I_{Rep}(p)=n \sum_{r=0}^R \sum_{k,l=r+1}^{R+1} \mu(r;k,l) 
{\rm Log}(p^2+\lambda(r;k,l)) }
Here the Replicon eigenvalue $\lambda$ is
\eqn\replic{ \lambda(r;k,l)=-2\tau -w {\hat q}_k -w {\hat q}_l -2 y q_r^2 }
The multiplicity $\mu(r;k,l)$ \DKT\ is given by 
\eqn\multip{ \mu(r;k,l)={1\over 2}(p_r-p_{r+1}) \mu(k)\mu(l) }
where 
\eqn\mumu{ \mu(k)=\left\{ \matrix{ {1\over p_k}-{1\over p_{k-1}}  & k>r+1\cr
{1\over p_{r+1}} & k=r+1\cr} \right. }
We note that $p_0$ is absent from $\hat q$, since even if the index
$k,l$ were allowed to take the value 0, it would appear in the vanishing
combination $p_0q_0$. The $p_0$ dependence can therefore only
arise from the multiplicity. Collecting the terms in $p_0$ we get
\eqn\newrep{ I_{Rep}={n\over 2} p_0 \sum_{k,l=1}^{R+1} \mu(k)\mu(l)
{\rm Log}(p^2+\lambda(0;k,l))
+{n \over 2}\sum_{r=1}^R p_r \sum_{k,l=r+1}^R{\rm Log}\left({p^2+\lambda(r;k,l)\over
p^2+\lambda(r-1;k,l)}\right) }
With the use of \exact-\twoeq\ and \fren-\mumu\ we get
\eqn\getlam{\eqalign{
\lambda(r;k,l)&=-2E(q_R)+g\big({q_R\over R}\big)^2({1\over 2}(k^2+l^2)-r^2-(k+l)+1)\cr
&=g q_R^2({1\over 2}(\big({k\over R}\big)^2+\big({l\over R}\big)^2)-\big({r\over R}\big)^2)
+g \big({q_R\over R}\big)^2(k+l-{5\over 6})\cr}}
We note that the lowest (Replicon) eigenvalue is given by 
\eqn\loev{ \lambda(r;r+1,r+1)= -{1\over 6}\big({q_R \over R}\big)^2 \qquad r=0,1,2,...,R}
hence we find an instability, except in the limit $R\to \infty$ where this is suppressed.
All other eigenvalues are positive. We thus have in the Parisi limit $R+1$ zero modes
arising from the negative eigenvalues \loev. In that limit, one has
\eqn\limparisi{ \eqalign{
I_{Rep}&= {n\over 2}\int_0^1 x dx \partial_x\left\{\int_x^1 {dk \over k} \partial_k \int_x^1 
{dl \over l}\partial_l {\rm Log}\big(p^2+g{q_R^2\over 2}(k^2+l^2-2x^2)\big)\right\}\cr
&+{n^2\over 2}\int_0^1{dk \over k}\partial_k \int_0^1{dl\over l} \partial_l {\rm
Log}\big(p^2+g{q_R^2\over 2}(k^2+l^2)\big) \cr}}
which coincides with (AM.12).

\newsec{Fluctuations: the longitudinal-anomalous (LA) sector}

We now have
\eqn\nowha{\eqalign{ n \delta f_{LA}&= {V\over 2} \int {d^Dp\over (2\pi)^D} \delta I_{LA}(p) \cr
\delta I_{LA}(p)&= n \sum_{k=0}^{R+1} \mu(k) {\rm Log} \det \Delta_k(r,s)\cr
\Delta_k(r,s)&=\delta_{r,s}^{Kr} -{w g q_{min(r,s)}\over \Lambda_k^{(r)}} \delta_s^{(k-1)} \cr}}
where $\delta^{Kr}$ denotes the Kronecker delta, while we have
\eqn\moredef{\eqalign{
\delta_s^{(k-1)} &\equiv p_s^{(k-1)}-p_{s+1}^{(k-1)}\cr
p_s^{(k-1)}&= \left\{ \matrix{ p_s & s\geq k-1 \cr 2 p_s & s<k-1 \cr}\right.\cr
\Lambda_k^{(r)}&= \left\{ \matrix{ p^2+\lambda(r;r+1,k) & k\geq r+1\cr
p^2+\lambda(r;r+1,r+1) & k< r+1\cr} \right.\cr}}
Expanding the determinant $\Delta_k(r,s)$ yields
\eqn\expandet{ \det \Delta_k(r,s)=1+\sum_{m=1}^\infty
(-w g)^n \sum_{0\leq s_1<s_2<...<s_m} \prod_{i=1}^m (q_{s_i}-q_{s_{i-1}}){\delta_{s_i}^{(k-1)}
\over \Lambda_k^{(s_i)}} }
where we have set $s_0\equiv 0$.
In order to have $p_0$ occurring in the determinant, i.e. in one of the $\delta_{s_i}^{(k-1)}$,
we need $s_i=0$ hence the only possible term is $s_1=0$, but the prefactor
$q_{s_1}-q_{s_0}\equiv q_{s_1}=q_0$ vanishes and there is no $p_0$ contribution
form the Log again. The only $p_0$ contribution comes from the multiplicity
$n\mu(k)$ which now cannot sustain an $n^2$ contribution\foot{One may then ask what becomes of the
term $k=0$ which has a factor $n/p_0$. It is actually given by
$\delta I_{LA}^{0} ={\rm Log} \det(\Delta_0(r,s)/\Delta_1(r,s))$, but with
$\delta_s^{(0)}=\delta_s^{(-1)}-p_0\delta_{s,0}^{Kr}$, whereas
$\Lambda_1(s)=\Lambda_0(s)$. Again, $\delta I_{LA}^{0}$ reduces to a contribution
$\sim p_0 \delta_{s,0}^{Kr}$ which vanishes with $q_0$.}.

\newsec{Conclusion}

With no contribution to fluctuations from the LA sector,
we conclude that the full answer is given by the contribution from
the Replicon sector as of 
\fren\ and \limparisi, thus corroborating the result of \AMOOR.

\newsec{Extension to Aspelmeier, Moore, Young calculation}

There the authors give an analytic answer to a long standing problem:
computing the interface free energy of the Ising spin glass. They show that (part of)
the answer is obtained by computing $\overline{Z^nZ^m}$ and, in 
the associated free energy, leaving aside
the terms in $(n+m)$, $(n+m)^2$ and keeping the terms in $nm$. From \newrep\ we clearly see that,
leaving aside a $(n+m)^2$ contribution, one is left with a term 
\eqn\perterm{ I_P(p)\equiv I_{Rep}(p)=-nm \sum_{k,l\geq 1}\mu(k)\mu(l)
{\rm Log}\big(p^2+\lambda(0;k,l)\big) }
replacing the quadratic term of \newrep\ and associated with periodic boundary conditions. 
The fluctuation contributions from the off-diagonal blocks of the Hessian
(mixed sector, associated with antiperiodic boundary conditions) is much easier to deal with
since it is {\it exactly} given by
\eqn\lutfin{I_{AP}(p)=+nm\sum_{k,l=0}^{R+1} \mu(k)\mu(l)
{\rm Log}\big(p^2+\lambda(0;k,l)\big)}
eigenvalues and multiplicities matching the $R=0$ calculation presented earlier
by the same authors \AMYSIX. The LA sector is represented here respectively by
$k=l=0$ and
$k=0$, $l\geq 1$ or $k\geq 1$, $l=0$, the Replicon sector by $k,l\geq 1$ as in \perterm.
But with ${\hat q}_1={\hat q}_0$ we now see that the two contributions
periodic \perterm\ and antiperiodic \lutfin\ are formally identical but for their sign.
This result again matches the one proposed in \AMYHUIT.

\noindent{\bf Acknowledgments} 

One of us (C.D.) is grateful to E. Brezin, M.A. Moore and A.P. Young
for enjoyable and useful discussions, and we both thank  T. Aspelmeier and M.A. Moore
for a clarifying exchange of correspondence.

\listrefs
\end